# Forecast-based charging strategy to prolong the lifetime of lithium-ion batteries in standalone PV battery systems in Sub-Saharan Africa

Jonathan Schulte [a,b,d]*, Jan Figgener [a,b,c]*, Philipp Woerner [a,b], Hendrik Broering [d], Dirk Uwe Sauer [a,b,c]

[a] Institute for Power Electronics and Electrical Drives (ISEA), RWTH Aachen University, Germany
[b] Institute for Power Generation and Storage Systems (PGS), E.ON ERC, RWTH Aachen University, Germany
[c] Juelich Aachen Research Alliance, JARA-Energy, Germany
[d] AMMP Technologies B.V., Netherlands
*These authors contributed equally to this paper,
Contact: jonathan.schulte@rwth-aachen.de

**Abstract -** Standalone PV battery systems have great potential to power the one billion people worldwide who lack access to electricity. Due to remoteness and poverty, durable and inexpensive systems are required for a broad range of applications. However, today's PV battery systems do not yet fully meet this requirement. Especially batteries still prove to be a hindrance, as they represent the most expensive and fastest-aging component in a PV battery system. This work aims to address this by prolonging battery life. For this purpose, a forecast-based charging strategy was developed. As lithium-ion batteries age slower in a low state of charge, the goal of the operation strategy is to only charge the battery as much as needed. The impact of the proposed charging strategy is examined in a case study using one year of historical data of 14 standalone systems in Nigeria. It was found that the proposed operation strategy could reduce the average battery state of charge by around 20% without causing power outages for the mini-grids. This would significantly extend the life of the battery and ultimately lead to a more durable and cheaper operation of standalone PV battery systems.

## 1. Introduction

In Sub-Saharan Africa, 548 million people in 2018 had no access to electricity [1]. The vast majority of them live in rural areas [2]. It is often impossible or economically unfeasible to connect these areas to the national energy grid [2]. Additionally, long-distance grid extensions are often unreliable, as 59% of rural households in Nigeria reported daily blackouts in 2016 [3]. To overcome the stated deficits, the World Bank identified that mini-grid solutions are essential and wants to connect 490 million people worldwide to a total of 210,000 mini-grids by 2030 [2]. For comparison, in 2019 47 million people worldwide were connected to 19,000, mini-grids [2]. Half of the planned mini-grids are announced to be built in Africa [2]. Historically, diesel generators have been a common technology for standalone energy systems in sub-Saharan

Africa [2]. A shift towards renewable solutions could be achieved through continuous cost reduction of batteries, photovoltaic (PV) panels, and PV inverters [2]. In addition, PV battery systems are more reliable, reduce air pollution, and are less noisy when compared to diesel generators [2, 4, 5].

Batteries contribute to a large part of the lifetime costs of PV battery systems [6, 7]. In addition to the already high initial cost, most of the batteries in this application have a comparatively short life of 5-10 years and therefore need to be replaced more frequently than PV panels or inverters, which nowadays can last up to 30 years. But due to difficult external conditions like high temperatures, sand, and torrential rain in Sub-Saharan Africa, a lifetime of 20 years is more realistic [2]. When replacing, not only the cost of the battery must be considered, but also the complicated procurement and installation in rural areas in Sub-Saharan Africa. Hence, it is desirable to maximize battery life.

For autonomous mini-grids in Sub-Saharan Africa, lithium-ion (Li-ion) batteries have overtaken lead-acid batteries and become the main battery technology [2]. In contrast to lead-acid batteries, Li-ion batteries should be operated at a low state of charge (SOC) to decelerate aging processes. Nevertheless, most solar mini-grids in Sub-Saharan Africa charge their battery whenever there is surplus energy available from solar generation. This can lead to batteries being operated in high SOCs for most of the time, which again accelerates battery aging.

### 1.1 Objective

This paper proposes a forecast-based operation strategy to extend the life of Li-ion batteries in standalone PV battery systems. The objective of the operation strategy is to only charge the battery as much as needed and thereby keep the battery in lower SOC. This dynamic operation is enabled by day-ahead forecasts of PV generation and consumption. To ensure that errors in the prediction don't lead to empty batteries, safeguards are introduced.

This work has access to one year of historical data from 14 standalone battery systems in Sub-Saharan Africa, which are used to train a forecast model as illustrated in Fig. 1. Based on this, the operation of the systems is simulated to investigate the impact of the proposed operational strategy on the 14 energy systems.

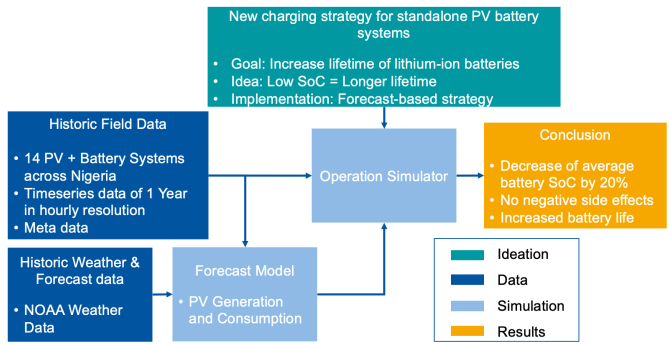

**Fig. 1.** *Overview of input data, ideation, and methodology*

This paper solely focuses on the reduction of the SOC to increase battery life. Other battery aging factors like temperature, charging rate, depth of discharge or cycle frequency are not optimized.

### 1.2 Literature review

Forecast-based charging strategies are established in other PV battery applications. Especially for grid-connected PV battery systems and microgrids, they have been researched for some time and are also used in commercial applications. The following overview of existing literature on forecast-based strategies in these applications is supported by a comprehensive list in Table 1.

For grid-connected PV battery systems, forecast-based charging strategies are mainly used to reduce the utility costs and increase the battery lifetime [8–12]. However, there are also strategies that neglect battery aging and only minimize utility costs [13, 14]. The latter is of no further interest for this work.

The former has been shown to prolong battery lifetime by 2-5 years on general lifetimes of 5-10 years [10–12]. The strategies aim to only charge as much energy into the battery as needed [8–12]. Thereby the SOC is kept relatively low, which in turn extends the calendar life of a Li-ion battery. Since full battery discharges are not a critical issue in grid-connected applications, the forecasts-based strategies here are designed in such a way that they can regularly lead to fully discharged batteries [10, 11].

In micro-grids, forecast-based charging strategies are used to optimize the interaction of PV, battery, diesel genset, and optionally wind and grid [15–20]. The objective function is to optimize the operation costs by reducing the costs of fuel, genset wear, and battery wear. Most available research focuses on lead-acid batteries, which have different aging mechanisms compared to Li-ion batteries.

The forecast-based charging strategy proposed in this work is mainly inspired by the approaches for grid-tied applications in [8–12]. In contrast to grid-tied systems, a full battery discharge leads to a power outage in standalone systems and consequently needs to be avoided. The proposed operation strategy addresses this risk by introducing safeguards.

### 1.3 Novelty and contribution

The selected references show potential for battery aging optimization in PV battery systems. However, all references focus on either grid-connected systems with a Li-ion battery or standalone systems with a lead-acid battery.
The novelty of this paper is to propose a battery life-extending operation strategy for standalone systems with Li-ion batteries.

Unlike most other work on operating strategies to extend battery life, this work has access to historical data from a large number of economically operated systems.

**Table 1**
Overview of existing literature on forecast-based charging strategies

| Title | Author | Application | Battery type | Battery aging considered? | Objective function | Forecast type |
|---|---|---|---|---|---|---|
| Aging-aware predictive control of PV-battery assets in buildings [8] | J. Cai, H. Zhang, X. Jin | Grid-connected PV battery system | Li-ion | Yes | Optimise utility costs and battery lifetime | Perfect forecast assumed |
| Optimal operation of hybrid PV-battery system considering grid scheduled blackouts and battery lifetime [9] | M. Alramlawi, A. Gabash, E. Mohagheghi, P. Li | Grid-connected PV battery system | Lead-Acid | Yes | Optimise utility costs and battery lifetime | Perfect forecast assumed |
| Comparison of different operation strategies for PV battery home storage systems including forecast-based operation strategies [10] | G. Angenendt, S. Zurmühlen, H. Axelsen, D.U. Sauer | Grid-connected PV battery system | Li-ion | Yes | Optimise utility costs and battery lifetime | Perfect forecast and persistence forecast |
| Enhancing Battery Lifetime in PV Battery Home Storage System Using Forecast Based Operating Strategies [11] | G. Angenendt, S. Zurmühlen, H. Mir-Montazeri, D. Magnor, D.U. Sauer | Grid-connected PV battery system | Li-ion | Yes | Optimise utility costs and battery lifetime | Perfect forecast and persistence forecast |
| The Impact of Control Strategies on the Performance and Profitability of Li-Ion Home Storage Systems [12] | N. Munzke, B. Schwarz, J. Barry | Grid-connected PV battery system | Li-ion | Yes | (Evaluating impact of predictive strategies on battery ageing) | Commercial Forecast |
| Operational Strategies for Battery Storage Systems in Low-voltage Distribution Grids to Limit the Feed-in Power of Roof-mounted Solar Power Systems [13] | A. Zeh, R. Witzmann | Grid-connected PV battery system | Li-ion | No | Optimise utility costs | Artificially created including error |
| Integration of PV Power and Load Forecasts into the Operation of Residential PV Battery Systems [14] | J. Weniger, J. Bergner, V. Quaschning | Grid-connected PV battery system | Li-ion | No | Optimise utility costs | Commercial and persistence forecast |
| Energy Management for Lifetime Extension of Energy Storage System in Micro-Grid Applications [15] | D. Tran, A. M. Khambadkone | Micro-grid (PV + Wind + Generator + Battery) | Li-ion and Lead-Acid | Yes | Optimize battery aging, power loss and power deviation | Markov-Chain |
| Assessing the value of forecast-based dispatch in the operation of off-grid rural microgrids [16] | S. Mazzola, C. Vergara, M. Astolfi | Micro-grid (PV + Genset + Battery) | Lead-Acid | Yes (only cyclic) | Optimize operation costs (fuel costs, genset wear, battery wear) | Perfect and erroneous forecast |
| A Model Predictive Control Approach to Microgrid Operation Optimization [17] | A, Parisio, E. Rikos, L. Glielmo | Micro-grid (PV + Grid + Battery + Genset) | Unknown | Yes (only cyclic) | Optimize operation costs (fuel costs, genset wear, battery wear, utility grid costs) | Support vector regression |
| Assessing the impact of a two-layer predictive dispatch algorithm on design and operation of off-grid hybrid microgrids [18] | L. Moretti, S. Polimeni, L. Meraldi, P. Raboni, S. Leva, G. Manzolini | Micro-grid (PV + Genset + Battery) | Li-ion and Lead-Acid | Yes (only cyclic) | Optimize operation costs (fuel costs, genset wear, battery wear) | SARIMA for Load forecast and artificially created for PV forecast |
| A Two-Stage Model Predictive Control Strategy for Economic Diesel-PV-Battery Island Microgrid Operation in Rural Areas [19] | J. Sachs, O. Sawodny | Micro-grid (PV + Genset + Battery) | Lead-Acid | Yes | Optimize operation costs (fuel costs, genset wear, battery wear) | SARIMA |
| Daily operation optimisation of hybrid stand-alone system by model predictive control considering ageing model [20] | R. Dufo-López, L. A. Fernández Jiménez, I. J. Ramírez-Rosado, J. S. Artal-Sevil, J. A. Domínguez-Navarro, J. L. Bernal-Agustín | Micro-grid (PV + Genset + Wind + Battery) | Lead-Acid | Yes (only cyclic) | Optimize operation costs (fuel costs, genset wear, battery wear) | Persistence Forecast |
| Control strategy for a standalone PV/battery hybrid system [21] | H. Mahmood, D. Michaelson, J. Jiang | Standalone PV battery | Lead-Acid | No | Only Control | No forecast used |
| Optimierung des Einsatzes von Blei-Säure-Akkumulatoren in Photovoltaik-Hybrid-Systemen unter spezieller Berücksichtigung der Batteriealterung [22] | D. U. Sauer | Standalone PV-Hybrid | Lead-Acid | Yes | Decelerate battery raging while ensuring robust electricity supply | No forecast used |
| Optimizing vehicle-to-grid charging strategies using genetic algorithms under the consideration of battery aging [23] | B. Lunz, H. Walz, D. U. Sauer | Battery in electric vehicle for V2G | Li-ion | Yes | Charge battery while optimising battery aging and energy trading profit | Perfect forecast assumed |

## 2. Methodology

This paper proposes a forecast-based operation strategy to increase the lifetime of Li-ion batteries in standalone PV battery systems. The goal of the proposed strategy is to only charge the battery as much as needed while minimizing the risk of an empty battery.

This paper has access to one year of historical data for 14 standalone PV battery systems in Sub-Saharan Africa monitored by the company AMMP. First, this data is examined to understand the historical operation and potential for optimization. Later, the historical data is used to simulate a scenario in which the systems are operated with the new proposed strategy.

Figure 2 gives an overview about the structure of this chapter, which starts with a brief description of the systems and the data set. Afterwards, the historical operation is analyzed with a focus on the potential for optimization. Further, the methods used to create PV generation and consumption forecasts are presented. Finally, the idea and algorithm of the proposed operation strategy are outlined.

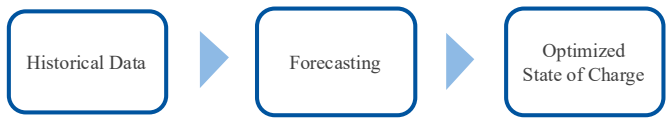

**Fig. 2.** Basic Idea of proposed operation strategy

### 2.1 Data Set

This section describes the application and topology of the PV battery system considered in this work. It also explains how the historical time series data was collected.

#### 2.1.1 Overview

All considered systems supply shops in a local market in Nigeria. This market is not connected to the national power grid. Before using PV battery systems, the market was powered by diesel generators.

**Table 2**
System overview

| | |
|---|---|
| *Number of systems* | 14 |
| *End-user* | Commercial |
| *Human settlement* | Urban |
| *Country* | Nigeria |
| *Battery chemistry* | Lithium-iron phosphate (LFP) |
| *Battery energy (per system)* | 10 kWh |
| *Installed PV power (per system)* | 9.75 kWp |
| *System coupling* | DC |

All considered systems share a similar typology, which is presented in

Fig. 3. The PV panels are DC coupled. Furthermore, lithium iron phosphate (LFP) batteries are used.

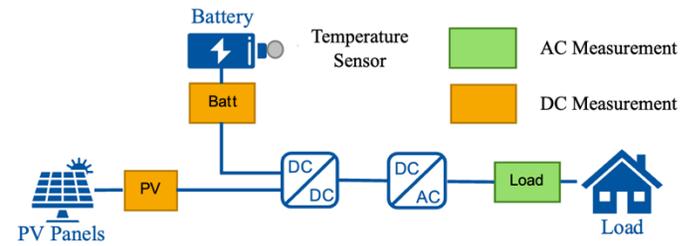

**Fig. 3.** System Topology and measurement point

At each site, two DC meters and one AC meter are installed, as illustrated in Fig. 3. The DC meters are integrated into the respective DC-DC converter to monitor the battery and the PV panels. For the AC meter at the load side, an external energy meter is used. Each meter measures voltage and current. Based on this, other quantities such as power, energy, and battery SOC are derived locally. Further, a temperature sensor is installed to monitor the battery temperature. Each device transmits its data to a controller device via the MODBUS protocol. From here, the data is transmitted either by a cellular or ethernet connection to a cloud database.

In this work, the following data is used in a 1-hour resolution:

- PV energy
- Consumption energy
- Battery SOC

#### 2.1.2 Historical operation

This section intends to give insights into the historical system operation to understand the further methodology. At first, a single system is analyzed in detail. This system is referenced as system 1. Afterwards, the other systems are included in the analysis.

The operation of system 1 is visualized for an exemplary week in Fig. 4. The upper part of the graph shows that the consumption (blue dotted line) is very similar on all days except Sunday. This is because the market is closed on Sundays. Moreover, there is no consumption at night because the stores are closed, and electricity consumption at night is even prohibited. The figure also displays the PV generation as a yellow dotted line. In the morning, PV generation often spikes before falling back to match consumption. To understand this behavior, it is helpful to look at the SOC curve. Here, it can be observed that the battery is fully charged every day already in the morning. As the system is not connected to a grid, the surplus PV power cannot be used to charge the battery and needs to be limited to the consumption demand. Therefore, the system generates a lot less energy than theoretically possible.

In the evening, the available PV power falls below consumption demand. Hence, the batteries are used to power the loads.

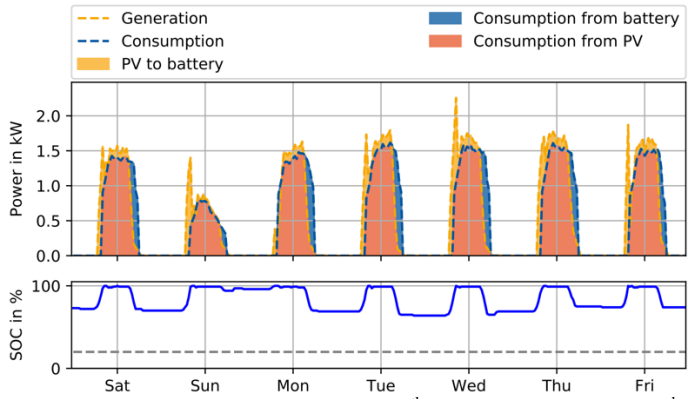

**Fig. 4.** Operation of system 1 from 16<sup>th</sup> November 2019 to 23<sup>rd</sup> November 2019

Figures 5 to 7 show the distribution of historical data for all recorded days of the given system to give a broader impression of the typical system operation. The data is presented as fan charts in such a way that each quarter hour of a day is assigned by the distribution of all values over the year in that quarter hour. Fig. 5 visualizes the consumption demand in which the median describes a typical commercial load profile. Figures 6 and 7 show the PV generation and battery SOC, respectively. Typically, at 5:00 AM, the PV panels start to generate electricity (Fig. 6). Because no consumption is required at that time, the energy is used to charge the battery. Therefore, the battery SOC (Fig. 7) rises between 5:00 AM and 7:00 AM. Between 7:00 AM and 8:00 AM in the morning, the SOC rises to 100%, and thus, the battery is fully charged.

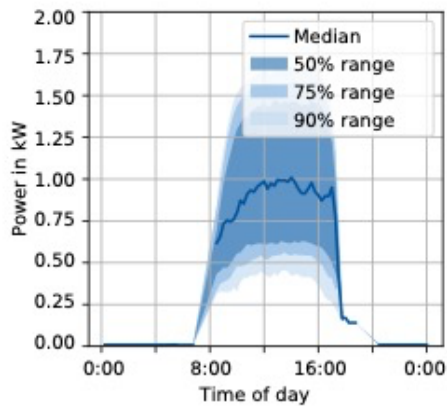

**Fig. 5.** Fan chart of Consumption (System 1)

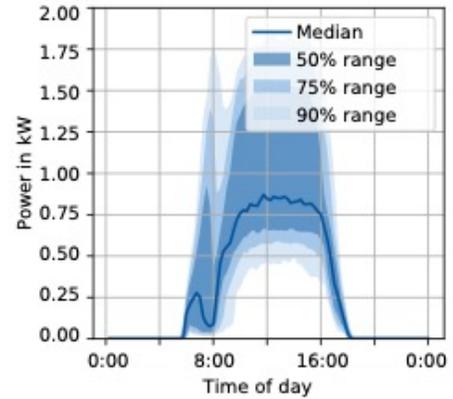

**Fig. 6.** Fan chart of PV Generation (System 1)

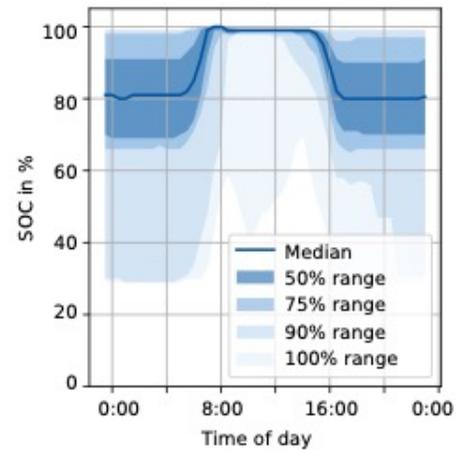

**Fig. 7.** Fan chart of battery SOC (System 1)

As soon as the battery is fully charged, the PV power (Fig. 6) is actively limited, as the PV power can only be used to meet the consumption demand. From here on, the PV generation follows the consumption curve for most of the day. Around 4 PM, the battery SOC starts to decline again. At this time, the PV power is no longer sufficient to meet the consumption demand. The battery steps in to supply the missing power. The battery reaches its minimum SOC around 5 PM and remains at the same level until the next morning.

The figures show that the battery is barely seeing any significant discharges and is kept fully charged for many hours of the day. Furthermore, the potential peak PV power of 9.75 kWp is not even closely reached on any day, as the fully charged battery leads to active curtailment of the PV power.

Table 3 shows various key performance indicators (KPIs) for each of the 14 systems. Each indicator is calculated for the whole period of 2019. The **PV generation** and **consumption** are close to each other because, with the exception of losses, only as much energy can be generated as is consumed locally in an off-grid application. The **average system efficiency** is calculated by dividing the consumed energy by the generated energy. The main reasons for losses in the systems are the efficiencies of inverter and battery as well as cable losses. The **75% SOC confidence interval** describes the window in which the 75% of the SOC values were located. This interval is formed around the median value. This KPI shows that the batteries of all systems are kept in a high or full SOC for most of the time. The **direct consumption rate** describes the proportion of the consumed electricity that came directly from the solar system without taking the detour via the battery. The **capacity factor** represents the ratio of the average PV power over the whole year and the rated peak power. A typical capacity factor for a PV system in Nigeria is 16% to 20% [24]. The values for the examined systems tend to be much lower, as the consumption of the systems is relatively low and the PV power is often actively curtailed as soon as the battery is fully charged. Finally, the **total downtime due to an empty battery** describes how often the battery of each system was fully discharged while there was insufficient PV power to power the loads. This finally leads to downtimes. It can be observed that downtimes tend to be more frequent for systems with high consumption.

With only one cycle per day and the batteries not being discharge below 60% - 80% on most day, the energy throughput of the battery is rather low. Further, the maximum charging / discharging power rates of 2 kW are rather low at 10 kWh battery capacity. While the rated maximal charge / discharge rate of the LFP batteries is around 1 C, the maxima in the operation of the analyzed systems is around 0.2 C. This leads to the assumption, that cyclic aging (influenced by charging frequency, charging rate, depth of discharge) has a minor role compared to calendar aging (mainly influenced by SOC and Temperature).

**Table 3**
Operational key indicators for each system in 2019

| System ID | PV Generation | Consumption | Average system efficiency | 75% SOC confidence interval | Direct consumption rate | Capacity factor | Total downtime due to empty battery |
|---|---|---|---|---|---|---|---|
| 1 | 2.99 MWh | 2.68 MWh | 89.6% | 77% - 100% | 94.4% | 3.5% | 0.0h |
| 2 | 2.56 MWh | 2.23 MWh | 87.1% | 91% - 100% | 96.3% | 3.0% | 0.0h |
| 3 | 2.61 MWh | 2.27 MWh | 86.9% | 92% - 100% | 96.4% | 3.0% | 0.0h |
| 4 | 2.67 MWh | 2.29 MWh | 85.8% | 92% - 100% | 97.0% | 3.1% | 0.0h |
| 5 | 2.96 MWh | 2.67 MWh | 90.2% | 84% - 100% | 92.2% | 3.5% | 3.5h |
| 6 | 7.27 MWh | 6.23 MWh | 85.7% | 50% - 100% | 91.0% | 8.5% | 126.0h |
| 7 | 4.62 MWh | 3.93 MWh | 85.1% | 82% - 100% | 94.0% | 5.4% | 0.0h |
| 8 | 4.41 MWh | 3.90 MWh | 88.5% | 65% - 100% | 95.0% | 5.2% | 48.0h |
| 9 | 3.78 MWh | 3.26 MWh | 86.2% | 66% - 100% | 89.0% | 4.4% | 21.0h |
| 10 | 4.07 MWh | 3.61 MWh | 88.7% | 70% - 100% | 94.0% | 4.8% | 25.5h |
| 11 | 3.31 MWh | 2.99 MWh | 90.3% | 83% - 100% | 95.0% | 3.9% | 4.5h |
| 12 | 2.36 MWh | 2.10 MWh | 88.9% | 90% - 100% | 96.5% | 2.7% | 0.0h |
| 13 | 2.61 MWh | 2.36 MWh | 90.0% | 90% - 100% | 96.0% | 3.1% | 10.5h |
| 14 | 3.9 MWh | 3.5 MWh | 89.9% | 79% - 100% | 91.0% | 4.6% | 0.0h |

## 2.2 Forecasts

The proposed operation strategy is based on forecasts. This chapter describes the algorithms used in this work to forecast both the day-ahead generation and consumption. Both forecasts are calculated in a resolution of 1h with a 24h time horizon.

### 2.2.1 Consumption Forecast

For the consumption forecast, a combination of a daily clustering technique and the autoregressive integrated moving average (ARIMA), as proposed in [25], is chosen. In contrast to [25], a two-level clustering method is used. At first, the weekdays are clustered based on the total energy consumption using the k-means Algorithm [26]. This typically led to one group of working days (Monday to Saturday) and one group for Sundays. The day for which the prediction is to be made is then associated with one of the clusters based on its day of the week. Finally, an ARIMA model is used to extrapolate the historical operation of the days in the respective cluster to the forecast day. Holidays are not considered in this work, which can increase accuracy even further [25]. It shall be noted that in contrast to [25], the number of clusters $k$ is not set automatically, but the optimal number of clusters is found as part of the algorithm using the silhouette method [26].

### 2.2.2 Generation Forecast

The goal of the generation forecast is to predict the theoretical PV generation power curve of the next day. Potential curtailments due to a fully charged battery are neglected. The basic idea of this forecast method is to use a horizontal irradiation forecast and convert it to a power forecast using a linear regression.

For the irradiation forecast, the publicly available historic downward short-wave radiation flux forecast from National Oceanic and Atmospheric Administration (NOAA) is used [27]. Since NOAA provides the data in a spatial resolution of 0.25 by 0.25 degrees, the irradiation forecast can be used accordingly for the individual locations.

To perform the linear regression, a linear least square regression model (LSR) is trained using historic irradiation data from NOAA (independent variable) and historic PV power (dependent variable) as proposed in [28].

To take the orientation of the PV panels into account, both the historic irradiation data and the irradiation forecasts are projected to the surface level of the PV panels with the help of the open-source tool *pvlib python* [29].

Real irradiation data from the sites for training the model or validating the data was not available.

Comparisons between the prediction and the real feed-in power are presented in Fig. 18 and Fig. 19.

### 2.3 Proposed operation strategy

In this section, a forecast-based charging strategy for standalone PV battery systems with a Li-ion is proposed. Firstly, the general objective and functioning of the strategy is outlined. Later, the exact algorithm of the proposed strategy is outlined.

### 2.3.1 Basic idea of proposed operation strategy

The calendric aging of a Li-ion battery is accelerated by high SOCs and high ambient temperatures [30–33]. The basic idea of the proposed algorithm is to keep the SOC as low as possible. Meanwhile, the SOC shall always stay above a certain security threshold to prevent the battery from being fully discharged during an unpredicted event.

Fig. 8 illustrates the basic idea of the presented algorithm by comparing it to a conventional strategy. Two approaches are used to decrease the SOC. While in the conventional operation of a PV battery system the battery is directly charged in the morning, the proposed strategy actively decides to delay the charging period to later in the day (similarly proposed in [12]). Hereby, the battery is kept longer at a low SOC. Additionally, an upper SOC cap is introduced to ensure that the battery is only charged to the maximum required SOC (similar to [11]).

Thereby, very high SOCs are avoided and the average SOC is decreased. The exact delay time and the exact SOC cap are dynamically calculated based on the forecasts. To prevent batteries from being fully discharged, two safety buffers have been built into the strategy. First, a conservative forecast is used instead of the mean forecast. Secondly, a buffer is kept free in the battery and is not used for the dynamic strategy.

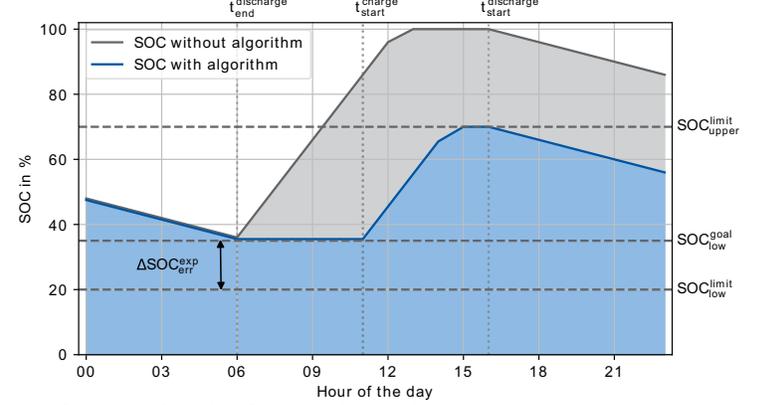

**Fig. 8**. Basic principle of proposed operation strategy

### 2.3.2 Algorithm of proposed operation strategy

The proposed operation strategy requires three input parameters, which are outlined in Table 4. The adjustable parameter $SOC_{low}^{limit}$ is used as a security buffer in case of an unforeseen event. $SOC_{low}^{limit}$ determines the portion of the battery charge that shall not be considered for the dynamic operation strategy. The algorithm will operate the battery in such a way that the SOC should not fall below this threshold in regular operation. If the SOC should fall below this threshold due to unforeseen events, the algorithm will try to charge the battery with any surplus power until the SOC is above $SOC_{low}^{limit}$ again. Through an amendment $SOC_{low}^{limit}$, the operation strategy can be tuned to be more conservative or more dynamic. The impact of such amendments will be examined in Section *III* of this paper. Further, the size of the battery $E_{batt}$ is required as input of the algorithm. Finally, $\eta_{charge}$ represents the energy efficiency from PV to battery and from battery to load. To reduce complexity, it is assumed that both are equal.

**Table 4**
List of input parameters

| Parameter | Definition |
|---|---|
| $SOC_{low}^{limit}$ | Adjustable security buffer |
| $E_{batt}$ | Size of battery in Wh |
| $\eta_{charge}$ | PV to battery (charge) and battery to load (discharge) efficiency (assumed to be equal) |

The algorithm defines three setpoints for the operation of the standalone system as outlined in Table 5. The exact value of each setpoint is dynamically calculated at each processing time $t_p$ of the algorithm. $SOC_{low}^{goal}$ represents the target value of the lowest SOC in case the forecasts are correct.

$SOC_{upper}^{limit}$ defines the upper SOC cap, while $t_{start}^{charge}$ determines the time at which to start charging the battery.

**Table 5**
Setpoints of operation strategy

| Setpoint | Definition |
|---|---|
| $SOC_{low}^{goal}$ | Target value of the lowest SOC of the day based on forecasted energy |
| $SOC_{up}^{limit}$ | Upper cap of SOC. PV power is curtailed if SOC should exceed this setpoint |
| $t_{start}^{charge}$ | Time at which to start charging the battery |

Before outlining the exact calculation of each of the setpoints, the forecast variables used in the algorithm are introduced in Table 6. The algorithm requires forecasts with a time horizon of 24h. While the upper index *"exp"* stands for the mean of the forecast, the upper indexes *"low"* and *"up"* index represent the lower and upper boundary of the 95% prediction interval.

**Table 6**
List of forecast variables

| Variable | Definition |
|---|---|
| $E_{pv}^{exp}(t_{start}, t_{end})$ | Expected PV energy generation between $t_{start}$ and $t_{end}$. *Equals* generation forecast. |
| $E_{pv}^{up}(t_{start}, t_{end})$ | Upper boundary of expected PV energy between $t_{start}$ and $t_{end}$ (95% prediction interval) |
| $E_{pv}^{low}(t_{start}, t_{end})$ | Lower boundary of expected PV energy between $t_{start}$ and $t_{end}$ (95% prediction interval) |
| $E_{cons}^{exp}(t_{start}, t_{end})$ | Expected consumption energy between $t_{start}$ and $t_{end}$. *Equals* consumption forecast. |
| $E_{cons}^{up}(t_{start}, t_{end})$ | Upper boundary of expected consumption energy between $t_{start}$ and $t_{end}$ (95% prediction interval) |
| $E_{cons}^{low}(t_{start}, t_{end})$ | Lower boundary of expected consumption energy between $t_{start}$ and $t_{end}$ (95% prediction interval) |

Using the energy forecasts, the algorithm predicts by how much the SOC will change in each of the considered time intervals. Here again, the median and the upper and lower quantiles are forecasted. By dividing the difference of the expected PV generation and the expected consumption by the energy capacity of the battery while multiplying it by the efficiency, the expected change of the SOC in each period is predicted (Eq. 1).

$$\Delta SOC^{exp}(t_{start}, t_{end})$$
$$= \frac{E_{pv}^{exp}(t_{start}, t_{end}) \cdot \eta_{charge} - E_{cons}^{exp}(t_{start}, t_{er})}{E_{batt}} \quad (1)$$

Similarly, the upper and lower boundaries of the prediction interval for the expected change of the SOC are calculated (Eq. 2, Eq. 3).

$$\Delta SOC^{up}(t_{start}, t_{end})$$
$$= \frac{E_{pv}^{up}(t_{start}, t_{end}) \cdot \eta_{charge} - E_{cons}^{low}(t_{start}, t_{end})}{E_{batt}} \quad (2)$$

$$\Delta SOC^{low}(t_{start}, t_{end})$$
$$= \frac{E_{pv}^{low}(t_{start}, t_{end}) \cdot \eta_{charge} - E_{cons}^{up}(t_{start}, t_{end})}{E_{batt}} \quad (3)$$

Based on the forecasts for the $\Delta SOC$, the algorithm calculates the setpoint $SOC_{low}^{goal}$. This setpoint is used as the target value for the lowest SOC of the next 24 hours. It is set in a way that the SOC stays above $SOC_{low}^{limit}$ even if the *"low"* scenario of the forecast occurs. $SOC_{low}^{goal}$ is determined by adding the expected error of the forecast during the discharge period to the security threshold $SOC_{low}^{limit}$ (Eq. 4).

$$SOC_{low}^{goal} = SOC_{low}^{limit}$$
$$+ \Delta SOC^{exp}(t_{sd}, t_{ed}) - \Delta SOC^{low}(t_{sd}, t_{ed}) \quad (4)$$

where:
$t_{sd}$ = start of next discharge period
$t_{ed}$ = end of next discharge period

After setting $SOC_{low}^{goal}$, the algorithm determines the setpoint $SOC_{up}^{limit}$. $SOC_{up}^{limit}$ is used as the upper charging cap. If the SOC should rise above $SOC_{up}^{limit}$, the PV power shall be curtailed to only power the consumption.

$SOC_{up}^{limit}$ is determined so that when the discharge period starts with that value, the battery will reach $SOC_{low}^{goal}$ at the end of the discharge period in case of the median forecast. Hence, $SOC_{up}^{limit}$ results from the sum of $SOC_{low}^{goal}$ and the expected $\Delta SOC$ of the discharge period (Eq. 5). The median forecast scenario (*"exp"*) is used in this step, as any considerations about inaccuracies of the forecast are already included in the calculation of $SOC_{low}^{goal}$.

$$SOC_{up}^{limit} = SOC_{low}^{goal} + \Delta SOC^{exp}(t_{sd}, t_{ed}) \quad (5)$$

Finally, the start of the charging period $t_{start}^{charge}$ is dynamically calculated by the algorithm. This setpoint allows delaying the charge of the battery system from the morning to later during the day. It is determined as the latest point in time at which starting to charge the battery would still be sufficient to reach $SOC_{up}^{limit}$ at the end of the charging period for the *"low"* scenario.

At each time $t_p$ the algorithm is processed, it questions whether the cumulative charge $\Delta SOC^{low}$ during the remaining charging period is sufficient to charge the battery up to $SOC_{up}^{limit}$ at the end of the charging period. This logic is reflected in Eq. 6. The operation strategy will start charging the battery once the term is true.

```
if:
```

$$SOC_{up}^{limit} \geq SOC(t_p) + \Delta SOC^{low}(t_p, t_{ec}) \qquad (6)$$

→ charge battery

```
else:
```

→ don't charge battery

```
where:
     tp = processing time
     tec = end of charge period
```

### 2.4 Simulation and assumptions

To evaluate the impact of the proposed forecast-based charging strategy, the operations under the new strategy for 2019 are simulated for each system and compared to real operations from that year. The simulation is based solely on energy flows and the SOC of the battery. To allow a calculation of the SOC purely based on energy quantities, it is assumed that the SOC is equal to the state of energy (SOE).

For the simulation of the load the real consumption data were used. The PV generation was often limited in real operation. For the simulation, it is necessary to consider how much PV energy could have been produced during these periods. For this purpose, historical irradiation data from NOAA were used. Based on that, the potential energy was calculated using the model of the solar plants of each system as in *C.2)*.

Efficiencies are respected in the simulation and are assumed to be static. For the LFP batteries, a DC-round-trip efficiency of 90% is considered [34]. For the inverter 94.3% [35] and for the MPPT 98% are assumed [36]. Any other losses, such as cable losses are neglected. The considered systems have a hard turn-off threshold if the battery SOC drops below 20%. If the SOC should drop below 20%, a power outage is assumed.

## 3. RESULTS AND DISCUSSION

The hypothesis of this paper is that the proposed forecast-based charging strategy can decrease the average battery SOC in standalone PV battery systems, without causing potential negative effects like a fully discharged battery causing a power outage. In this work, this hypothesis is tested using a simulation-based approach. To evaluate the consequences of the new operating strategy, the reduction in average SOC and the reduction in the time the battery is fully charged are compared with the increase in power outages. All these indicators are measured over the course of an entire year.

The proposed operation strategy can be set as more dynamic or conservative. This is done by increasing/decreasing the parameter $SOC_{low}^{limit}$. As described in section *II.D.*, the parameter describes the portion of the battery charge that is not included in the dynamic calculations but is kept as a buffer. A lower value means less buffer and thereby a more dynamic strategy. The comprehensive impact of setting different values for this parameter are outlined later in this chapter.

Firstly, a closer look at the impact of the new strategy on the daily operation based on the examples of two systems over a short period is taken. For those examples, $SOC_{low}^{limit} = 60\%$ is used. Afterwards, the results of choosing different values for $SOC_{low}^{limit}$ are analyzed in detail for one system over a period of a year. Lastly, an overview of the full-year analysis for different $SOC_{low}^{limit}$ and all 14 systems is presented.

Fig. 9 shows the operation of system 1 for the old and the new strategy in a period from 2nd November 2019 to 6th November 2019. The upper graph shows the power balance of the old operation and the middle graph for the new operation. The lower graph displays the SOC of both the old and new strategies. The results show that the new operation strategy leads to less battery charging in the morning. Especially on the 3rd November 2019, when the battery is not charged at all in the morning. Furthermore, the new SOC curve shows that the battery is not fully charged at the end of the day anymore. Thereby, the SOC can be held equal to or lower than the real SOC over the whole period. This period represents a successful example of the new operation strategy, as the average SOC can be reduced while the battery SOC never drops below 20%, and thereby no power outages are caused.

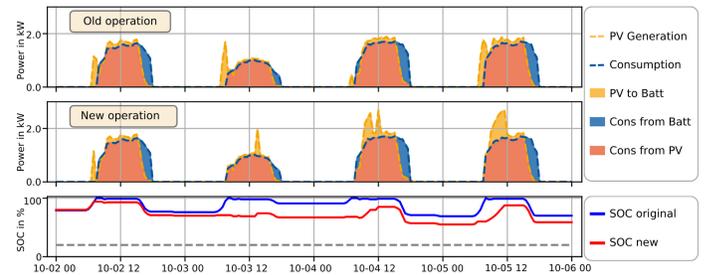

**Fig. 9.** Comparison of old and new operation strategy for system 1 from 2nd November 2019 to 6th November 2019; $SOC_{low}^{limit} = 60\%$

Fig. 10 illustrates a similar graphic for system 7. Compared to system 1, system 7 has a much higher average consumption. For system 1 the dynamic operation strategy does not always fully charge the battery. But as the algorithm recognizes that the consumption is much higher for system 7, the algorithm decides to charge the battery fully and thereby mitigate the risk of a dead battery. The only benefit of the new operation strategy for this system is that the charging period can be postponed from morning to noon or afternoon. Thereby the time in which the battery is fully charged is reduced by around 50%. The impact of the security threshold $SOC_{low}^{limit}$ can be well observed in this example. On both 4$^{th}$ October 2019 and 5$^{th}$ October 2019, the battery SOC is below the set threshold of $SOC_{low}^{limit} = 60\%$. Hence the battery is immediately charged up to 60% in the morning. Afterwards, the charging process is stopped for a few hours. Finally, during midday, the battery gets fully charged.

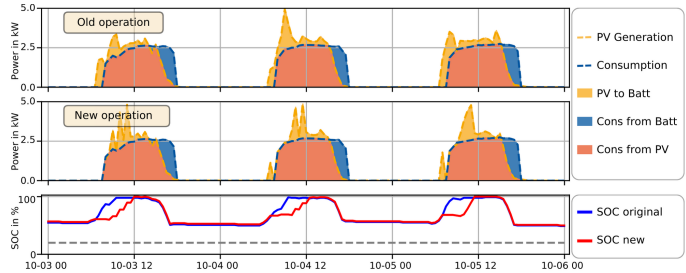

**Fig. 10.** Comparison of old and new operation strategy for system 7 from 3$^{rd}$ October 2019 to 6$^{th}$ October 2019; $SOC_{low}^{limit} = 60\%$

After a closer look at examples of operation over a few days, the operation of system 1 over a whole year and for different values of $SOC_{low}^{limit}$ will now be analyzed. Figures 11 to 14 illustrate the SOC distribution of the old and new strategies with different values for $SOC_{low}^{limit}$. Fig. 11 presents the SOC distribution for the old static operation strategy. It can be observed that the battery is at a high or full SOC for most of the time and is never fully discharged. Figures 12 to 14 show the resulting SOC distribution for the proposed dynamic strategy using different values for the parameter $SOC_{low}^{limit}$. The figures show that a lower $SOC_{low}^{limit}$ generally leads to a lower SOCs. The SOC distribution of the conservative strategy with an $SOC_{low}^{limit} = 80\%$ is similar to the previous static operation. A difference can be observed in the morning. Here, the battery is fully charged less frequently because the dynamic operation strategy delays the charging. This results in an average SOC of 82.2% while it was 85.2% with the old static operation strategy. Another indicator for the decrease of the battery SOC is the average time per day when the battery is fully charged. This time could be reduced from 11.5 hours per day to 7.1 hours per day.

For the conservative strategy and the old strategy, no power outages have occurred. For the moderate strategy with $SOC_{low}^{limit} = 60\%$, the average SOC decreased to 71.2%. The average fully charged hours per day are reduced to 2.4 hours a day. However, the battery is completely discharged once, which would have led to a power failure of 2 hours. In the old operation strategy, 0 hours of power outages due to an empty battery have occurred. The most dynamic strategy with $SOC_{low}^{limit} = 40\%$ is reducing the SOC even further with an average SOC of 57.6% and an average of 2.1 hours a day when the battery is fully charged. As a negative result, the power outages cumulate to 14.5 hours in the considered year.

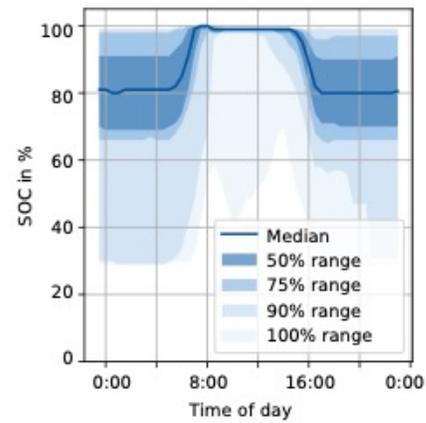

**Fig. 11.** SOC range with old operation strategy for system 1

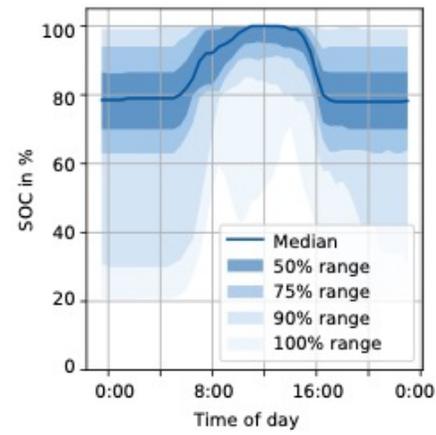

**Fig. 12.** SOC range with conservative strategy: $SOC_{low}^{limit} = 80\%$

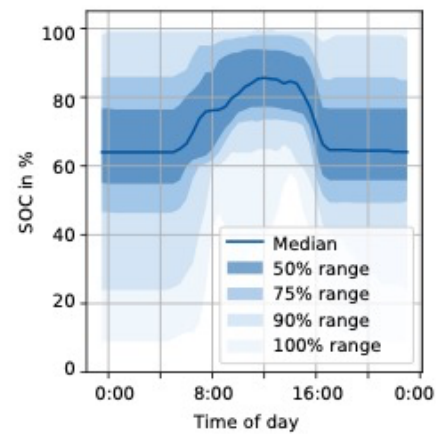

**Fig. 13.** SOC range with moderate strategy: $SOC_{low}^{limit} = 60\%$

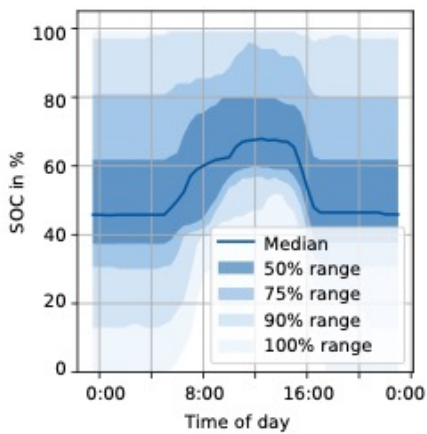

**Fig. 14.** *SOC range with dynamic strategy:* $SOC_{low}^{limit} = 40\%$

Finally, the impact of the proposed strategy for all systems is evaluated. Therefore, the full simulation is performed for each system and for 40 different $SOC_{low}^{limit}$ values between 20% and 100%. In each iteration, the following key indicators are calculated:

- Total power outage time due to empty battery
- Average SOC
- Average time per day when the battery is fully charged

Fig. 15 visualizes the power outage times for the new operation strategies for all 14 systems. The x-axis displays the respective value $SOC_{low}^{limit}$. As a reference, the power outage times for the old operation strategy are plotted for each system on the right side of the graphic.

At $SOC_{low}^{limit} = 100\%$, the new operation strategy achieves the same result as the old strategy. This is reasonable, as the full battery capacity is used as a security buffer. Thus, the battery is always charged, when possible, which is exactly the approach of the old operation strategy. For $65\% < SOC_{low}^{limit} < 100\%$, the power outage times barely increase. The SOC is held sufficiently high to avoid additional outages. It shall be noted that the value of 65% is just an empirical finding for this data sample. When $SOC_{low}^{limit}$ is lower than 65%, the occurrences of power outages increase almost consistently for all systems. Choosing a $SOC_{low}^{limit}$ in this range can still be economically most beneficial. However, a comprehensive trade-off between the costs of power outages and the benefits of decelerated battery aging must be carried out.

In general, the graphic shows that the new operation strategy with $SOC_{low}^{limit} = 65\%$ would have resulted in a safe operation without significant increases in power outages. The benefits for battery aging when using this strategy are outlined based on Fig. 16 and Fig. 17.

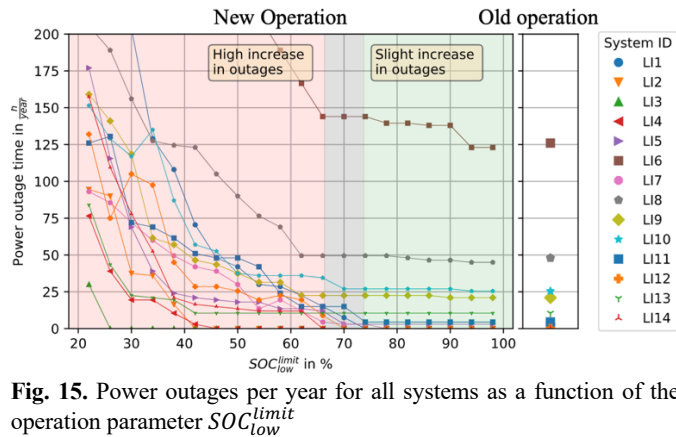

**Fig. 15.** Power outages per year for all systems as a function of the operation parameter $SOC_{low}^{limit}$

Fig. 16 outlines the impact of the new operation strategy on the average SOC. The average SOC serves as an indicator of whether the algorithm can reduce the SOC significantly. As outlined before, the result of the new operation strategy for $SOC_{low}^{limit} = 100\%$ is identical to the old strategy. For $SOC_{low}^{limit}$ between 80% and 100%, the curves have a rather low gradient. In this range, only the charging delay has an impact on the operation and can slightly decrease the average SOC. Starting from $SOC_{low}^{limit} < 80\%$, the dynamic algorithm additionally also decides to stop charging the battery at an upper charging threshold more frequently, which leads to a steeper decrease of the average SOC. A dotted line at $SOC_{low}^{limit} = 65\%$ is drawn, as it was shown before that no additional power outages are expected for $SOC_{low}^{limit} > 65\%$.

When comparing the old operation strategy with the new operation strategy at $SOC_{low}^{limit} = 65\%$, the figure shows that the average $SOC_{low}^{limit}$ is reduced for all systems. For the old operation strategy, the average SOC across all systems is 88.5%. For the new operation strategy at $SOC_{low}^{limit} = 65\%$, the average SOC across all systems is brought down to 72%.

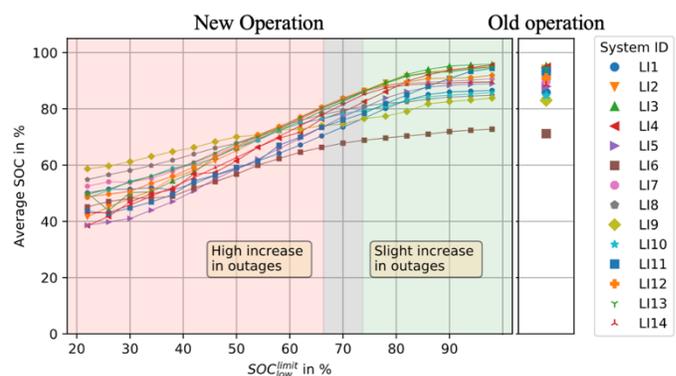

**Fig. 16.** Average SOC for all systems as a function of the operation parameter $SOC_{low}^{limit}$

Fig. 17 visualizes the average time per day in which the battery is fully charged, again for different values of $SOC_{low}^{limit}$. This indicator is interesting for the evaluation of the algorithm, as the aging of a Li-ion battery progresses at the fastest pace if the battery is fully charged.

For $80\% < SOC_{low}^{limit} < 100\%$, the curves of most systems show a steep slope. Even though the batteries are still getting fully charged here, the charging delay reduces the time they are fully charged significantly. At around $60\% < SOC_{low}^{limit} < 80\%$, the algorithm decides to decrease the upper charging cap and stops fully charging the battery more often. For $SOC_{low}^{limit} < 60\%$, the systems reach saturation for fully charged time per day between 0 and 2 hours a day.

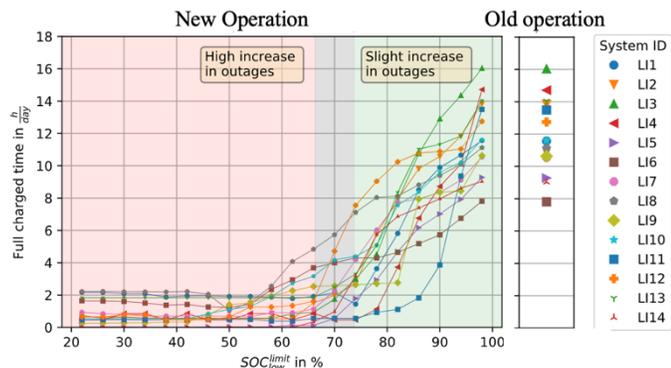

**Fig. 17.** Average full charged time per day for all systems as a function of the operation parameter $SOC_{low}^{limit}$

In summary, the results show that it is possible to decrease the average SOC by ~20% and the average full charged time per day for all systems by ~7 hours without causing a fully discharged battery leading to significantly more power outages. Various sources have shown that the reduction of the average SOC and the average fully charged time, leads to a longer lifetime of Li-ion batteries [30–33].

The results from the other literature evaluating similar operating strategies [10,11], but for grid-connected systems, showed a reduction of the average SOC by 13%. The comparatively higher reduction in this work can be explained by the fact that the systems considered in this work were operated at a high average SOC of 70%-90% with the old operation strategy, while the grid-connected system in [10,11] had an average SOC of 30%. Thus, there is more room for SOC reduction for the systems considered in this paper. It is worth noting, that standalone systems tend to have higher average SOCs when compared to grid-connected systems since the battery must not be fully discharged. Therefore, larger sizing leads to a higher buffer.

Using realistic assumptions, true field data, and real forecasts in this work, the results promise high validity. However, as the simulation is only based on energy flows, a more complex simulation or practical field tests would lead to higher plausibility.

## 4. Conclusion

Rural electrification, climate change goals, and decreasing costs of PV systems and batteries enable the PV off-grid solution to gain significance in the African energy market. However, the market is very cost-sensitive and requires durable solutions. As batteries are the most expensive and least durable component of a standalone PV battery system, this work proposes a forecast-based charging strategy to extend battery life.

As the aging of lithium-ion batteries is accelerated when they are operated in a high state of charges, the proposed strategy pursues the goal of only charging the battery as much as necessary. To mitigate the risk of a dead battery and potential power outages, various safeguards are added to the algorithm.

The reduction of the SOC is accomplished by two approaches. On one hand, the charge of the battery is postponed from the morning to the afternoon. On the other hand, an upper SOC cap, above which the battery is stopped being charged, is introduced. With the help of PV generation and consumption forecasts, the algorithm decides to which extent each of the approaches shall be used. As forecasts are not always accurate, an adjustable buffer is used as a safeguard.

To evaluate the impact of the proposed strategy, a case study with 14 standalone PV battery systems powering a local market in Nigeria was conducted. Simulating the operation of considered systems for one year showed that the proposed charging strategy can be beneficial. Without causing additional power outages, the new strategy reduces the average SOC by ≈ 20% for most systems while not causing additional power outages. Similarly, the average time per day in which the battery is fully charged was reduced by ≈ 7 hours. When tuning the strategy more dynamic, the SOC can be reduced even further. However, this comes at the cost that the battery is more often fully discharged which leads to power outages. Further research should investigate the period by which lowering the SOC through this operating strategy extends the life of an LFP battery.

The results of this work can be extended to any standalone PV battery system and are not geographically limited. Systems with high average SOCs and low utilization of the battery can profit most, as the average SOC can be reduced to a greater extent, which in turn leads to a higher battery lifetime. Finally, this new forecast-based operation strategy may be of particular interest for batteries in tropical climates, as fast battery degradation poses a greater problem there."

## Acknowledgement

We thank AMMP Technologies B.V. (URL: https://www.ammp.io/) for providing the data set and assistance.

## Author Contributions

**Jonathan Schulte**: Conceptualization, Methodology, Software, Formal analysis, Investigation, Data Curation, Writing – Original Draft, Visualization, Validation. **Jan Figgener**: Conceptualization, Methodology, Software, Formal analysis, Investigation, Writing – Original Draft, Visualization, Validation. **Philipp Woerner**: Writing – Original Draft, Visualization. **Hendrik Broering**: Conceptualization, Methodology, Data Curation, Writing – Review & Editing. **Dirk Uwe Sauer**: Conceptualization, Methodology, Resources, Writing – Review & Editing, Supervision.

## Appendix

**Table 7**
List of acronyms

| Abbreviation | Definition |
| --- | --- |
| ARIMA | Autoregressive integrated moving average |
| KPI | key performance indicator |
| Li-ion | lithium-ion |
| LFP | lithium iron phosphate |
| LSR | linear least square regression model |
| NOAA | National Oceanic and Atmospheric Administration |
| PV | photovoltaic |
| SOC | state of charge |
| SOE | state of energy |

**Table 8**
List of symbols

| Symbol | Unit | Definition |
| --- | --- | --- |
| $E_{batt}$ | Wh | Size of battery in Wh |
| $E_{pv}^{exp}(t_{start}, t_{end})$ | Wh | Expected PV energy generation between $t_{start}$ and $t_{end}$ |
| $E_{pv}^{up}(t_{start}, t_{end})$ | Wh | Upper boundary of expected PV energy between $t_{start}$ and $t_{end}$ (95% prediction interval) |
| $E_{pv}^{low}(t_{start}, t_{end})$ | Wh | Lower boundary of expected PV energy between $t_{start}$ and $t_{end}$ (95% prediction interval) |
| $E_{cons}^{exp}(t_{start}, t_{end})$ | Wh | Expected consumption energy between $t_{start}$ and $t_{end}$ |
| $E_{cons}^{up}(t_{start}, t_{end})$ | Wh | Upper boundary of expected consumption energy between $t_{start}$ and $t_{end}$ (95% prediction interval) |
| $E_{cons}^{low}(t_{start}, t_{end})$ | Wh | Lower boundary of expected consumption energy between $t_{start}$ and $t_{end}$ (95% prediction interval) |
| $\eta_{charge}$ | % | PV to battery and battery to load efficiency (assumed to be equal) |
| $SOC_{low}^{limit}$ | % | Adjustable security buffer |
| $SOC_{low}^{goal}$ | % | Target value of the lowest SOC of the day based on forecasted energy |
| $SOC_{up}^{limit}$ | % | Upper cap of SOC. PV power is curtailed if SOC should exceed this setpoint |
| $\Delta SOC^{exp}(t_{start}, t_{end})$ | % | Expected change of SOC between $t_{start}$ and $t_{end}$ |
| $\Delta SOC^{up}(t_{start}, t_{end})$ | % | Upper boundary of expected change of SOC between $t_{start}$ and $t_{end}$ |
| $\Delta SOC^{low}(t_{start}, t_{end})$ | % | Lower boundary of expected change of SOC between $t_{start}$ and $t_{end}$ |
| $t_{start}^{charge}$ | s | Time at which to start charging the battery |
| $t_{sd}$ | s | Start of next discharge period |
| $t_{ed}$ | s | End of next discharge period |
| $t_p$ | s | Time of processing |
| $t_{ec}$ | s | End of charge period |

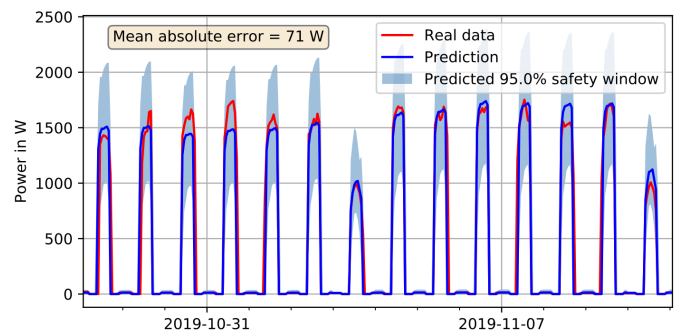

**Fig. 18**: Exemplary result of consumption forecast for system 1

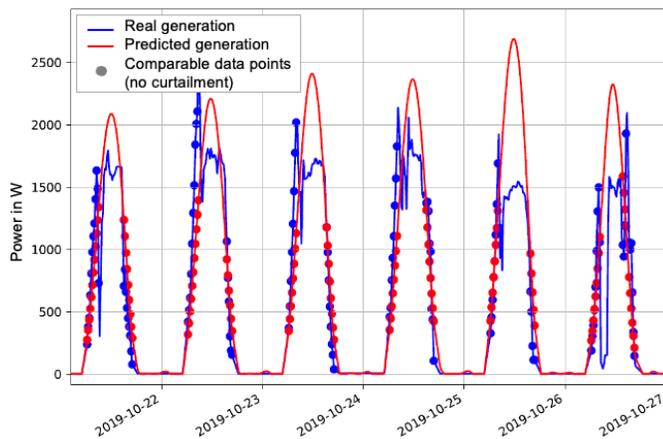

**Fig. 19:** *Exemplary result of generation forecast for system 1 - only marked points are comparable, as real PV generation is mostly curtailed at midday due to fully charged battery*